# Binding energy referencing in X-ray photoelectron spectroscopy: expanded data set confirms that adventitious carbon aligns to the vacuum level

Grzegorz Greczynski

*Thin Film Physics Division, Department of Physics (IFM), Linköping University, SE-581 83 Linköping, Sweden*

**Abstract**

The correct referencing of the binding energy (BE) scale is essential for the accuracy of chemical analysis by *x*-ray photoelectron spectroscopy. The C 1s C-C/C-H peak from adventitious carbon (AdC), most commonly used for that purpose, was previously shown to shift by several eVs following changes in the sample work function $\phi_{SA}$, thus indicating that AdC aligns to the sample vacuum level (VL). Here, results from a much larger sample set are presented including 360 specimens of thin-film samples comprising metals, nitrides, carbides, borides, oxides, carbonitrides, and oxynitrides. Irrespective of the material system the C 1s peak of AdC is found to follow changes in $\phi_{SA}$ fully confirming previous results. Several observations exclude differential charging as plausible explanation. All experimental evidence points instead to the VL alignment at the AdC/sample interface as the source of shifts. Should the C 1s peak of AdC be used for spectra referencing the measurement of sample work function is necessary, irrespective of whether samples are measured grounded or insulated from the spectrometer.

Reliable binding energy referencing is crucial in XPS as it ensures that peaks appear at correct values so that by comparing to literature or data bases assignments to chemical states can be made. To be able to make reliable chemical bonding assessments, the instruments' energy scale needs to be correctly calibrated,[1,2] and a reliable energy reference from the sample of interest must be available. The former procedure, performed on a regular basis, ensures that the instrument works correctly, i.e., that the peaks from standard samples (typically sputter-etched metals such as Au, Ag, and Cu) appear at the correct positions on the binding energy scale.[1,3] This, however, does not guarantee that the same holds for signals from any other sample outside the calibration set. The reason is the occurrence of sample charging, which, with the exception for metals making good contact to the spectrometer, is an unknown quantity. For that reason, the reference signal from the sample itself is required if extracted $E_B$ values should have an absolute character.

By far the most common way to charge reference XPS spectra is to use the $E_B$ of the C-C/C-H component of the C 1s spectrum from adventitious carbon (AdC) layer present on majority of samples.[4] C 1s peak is then set at the arbitrary chosen value from the range 284.6-285.0 eV.[5,6] The method was most frequently criticized in 70's and 80's for the unknown nature of AdC, uncertain electrical equilibrium with the surface it accumulates on, and the C 1s $E_B$ varying with the AdC thickness. [7,8,9,10,11,12,13,14,15] Despite that none of the objections were disproved, the method has gained tremendously in popularity since then, predominantly due to its simplicity. More structured criticism came during recent years from our lab, based on the systematic studies of thin film samples deposited on conducting substrates covering different materials systems.[16] We reported that the AdC signature depends on the substrate type, exposure time, and the environment,[17,18] which confirms the earlier doubts.

In the follow up paper,[19] we analyzed the BE of the C 1s peak from nearly a hundred AdC layers accumulating on a whole range of substrates including metals, nitrides, carbides, oxides, borides. The $E_B^F$ of the C-C/C-H peak was found to vary over an alarmingly large range, from 284.08 to 286.74 eV, depending on the substrate, clearly disproving the common assumptions. Irrespectively of the materials system studied, the C 1s $E_B^F$ correlated to the sample work function $\phi_{SA}$, such that the sum $E_B^F + \phi_{SA}$, corresponding to the binding energy referenced to the vacuum level $E_B^V$, was nearly constant at 289.58±0.14 eV. The latter indicates that AdC aligns to the sample vacuum-level (VL),[19] which, as established in the early days of XPS,[20,21,22] implies that a complementary measurement of $\phi_{SA}$ has to be performed if the C 1s peak of AdC should serve the purpose of binding-energy scale referencing. The C 1s BE should then be set at 289.58 - $\phi_{SA}$ eV, and all other core levels should be aligned accordingly.[19]

Here, previous finding of the VL alignment at the AdC/sample interface is tested for a much broader sample set comprising 360 thin-film specimens spanning a wide range of material systems such as metals, nitrides, carbides, borides, oxides, carbonitrides, and oxynitrides. The list of samples is included in the supplementary file. The samples are grown by magnetron sputtering methods from elemental as well as compound targets, operated either in Ar or in ($Ar/N_2$) gas mixture. Substrates include Si(001), $Al_2O_3$(111), and steel. Experiments are performed in various vacuum systems with widely varying background pressures that range from ultra-high vacuum (UHV) to high-vacuum conditions. The growth temperature $T_s$ is from RT (23 °C) to 900 °C, while the venting temperature $T_v$ (temperature at which samples are first exposed to the ambient)[23] ranges from RT to 330 °C. Film thicknesses determined from cross-sectional scanning electron microscopy analyses range from 2 to 2830 nm. Samples have been exposed to the ambient environment for time periods ranging from several weeks to a few years.

All analyzes are performed in the same spectrometer (Axis Ultra DLD from Kratos Analytical) employing monochromatic Al Kα radiation source (hν = 1486.70 eV). The base pressure during analyses is better than $1.1\times10^{-9}$ Torr ($1.5\times10^{-7}$ Pa). Immediately after XPS analyses the sample work function measurements are conducted by ultraviolet photoelectron spectroscopy (UPS) with unmonochromatized He I radiation (hν = 21.22 eV) and negatively biased samples. $\phi_{SA}$ is assessed from the secondary-electron cutoff energy in the He I UPS spectra by a linear extrapolation of the low-kinetic-energy electron tail towards the BE axis,[24,25] with a precision of ±0.05 eV. $\phi_{SA}$ obtained from the sputter-cleaned reference Au sample is 5.13 eV, in very good agreement with the textbook values that range from 5.0 to 5.4 eV.[24] The calibration of the binding-energy scale is confirmed by examining sputter-cleaned Au, Ag, and Cu samples.[1,3] The charge neutralizer was not used in any of the reported experiments. For all samples that exhibit the Fermi level cut-off XPS valence band (VB) spectra are acquired to confirm whether the FE is at 0 eV on the BE scale, thus proving that the samples are in good electrical contact to the spectrometer.

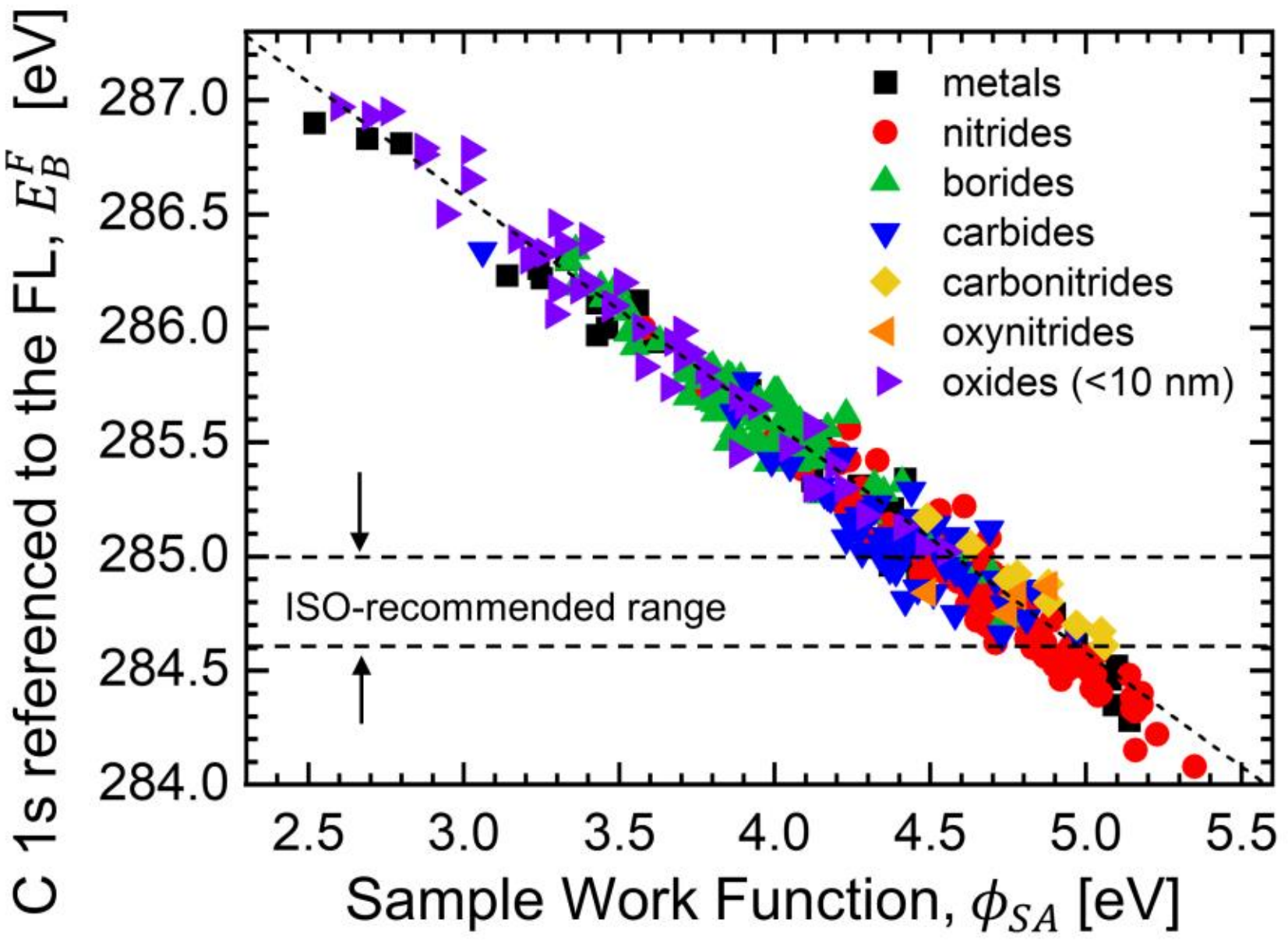

**Figure 1** (color online) Binding energy of the C 1s peak corresponding to C-C/C-H bonds of adventitious carbon referenced to the Fermi level $E_B^F$ plotted as a function of sample work function $\phi_{SA}$ assessed by UPS. 360 thin-film specimens spanning a wide range of material systems such as metals, nitrides, carbides, borides, oxides, carbonitrides, and oxynitrides are included. All samples are in the form of thin films and have been exposed to the ambient conditions for a time period ranging between several weeks and a few years.

In Fig. 1 the position of the C 1s C-C/C-H peak of AdC measured with respect to the spectrometers' FL $E_B^F$ is plotted as a function of the sample work function $\phi_{SA}$ for more than 360 samples of metals, nitrides, borides, carbides, carbonitrides, oxynitrides, and oxides. The very close correlation between $E_B^F$ and $\phi_{SA}$ is evident. In fact, the sum $E_B^F + \phi_{SA}$ is equal to 289.58±0.12 eV. The average value agrees perfectly with previous results obtained for a limited sample set,[19] while the standard deviation is reduced from 0.14 to 0.12 eV. The variation in the C 1s peak position is 2.89 eV, which is seven times more than the range specified in ISO guidelines, and more than many chemical shifts. Among samples showing the highest $E_B^F$ of the C 1s peak are several metals such as Mg (286.90 eV, $\phi_{SA}$ = 2.52 eV), Mg-Al alloy (286.83 eV, $\phi_{SA}$ = 2.69 eV), and Y (286.81 eV, $\phi_{SA}$ = 2.8 eV). Samples with lowest BE of the C 1s peak are several nitrides: MoN (284.08 eV, $\phi_{SA}$ = 5.35 eV), WN (284.22 eV, $\phi_{SA}$ = 5.23 eV), and VN (284.15 eV, $\phi_{SA}$ = 5.16 eV). Such large C 1s shifts can be one possible reason for the large spread of BE values reported for the same chemical state in XPS data bases, e.g., 2.5 eV for the Zr $3d_{5/2}$ peak of $ZrO_2$, 2.3 eV for the Al 2p peak of $Al_2O_3$, 1.6 eV for the Na 1s peak of NaCl, 1.5 and 1.1 eV for the Si 2p peak from $Si_3N_4$ and $SiO_2$.[26]

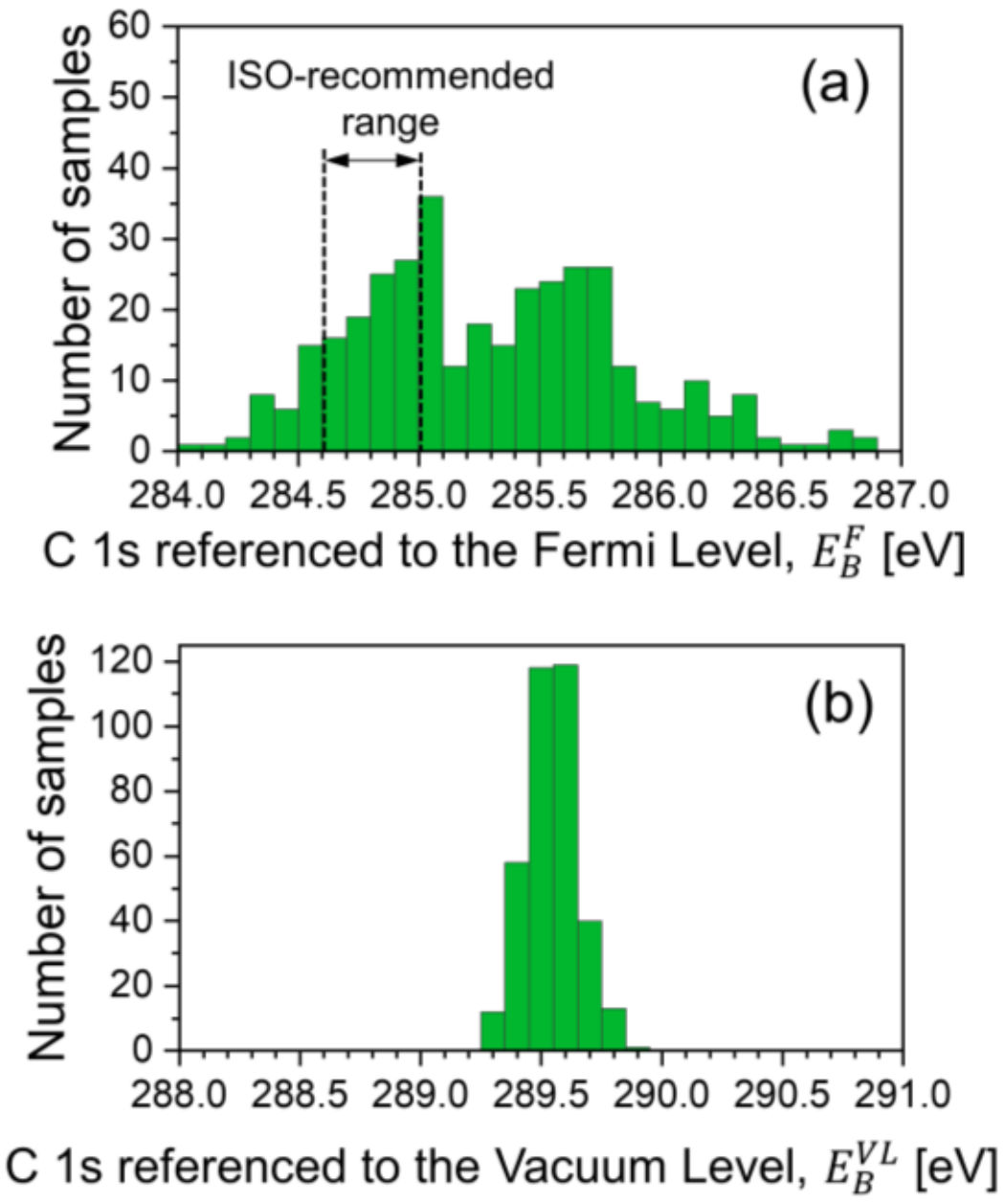

**Figure 2** (color online) Histograms showing the distribution of the C 1s peak position referenced to (a) the Fermi level $E_B^F$, and (b) the Vacuum level $E_B^{VL}$. Plots illustrate differences between the conventional FL referencing and the VL referencing that involves the measurement of the sample work function. 360 thin-film specimens spanning a wide range of material systems such as metals, nitrides, carbides, borides, oxides, carbonitrides, and oxynitrides are included.

The subset of ultrathin oxide films is also added to Fig. 1. Such thin layers, insulating by nature, were recently shown to align to the VL provided that the oxide thickness is comparable to the electron penetration range such that the severe charging effects are prevented by the substrate electrons.[27] Up to the thickness of 12 nm the core level peaks from oxide (e.g. Al 2p from $Al_2O_3$) followed changes in the sample work function in the same manner as the C 1s peak of AdC. Hence, the VL alignment appears to be a general condition for thin layers with insulating properties (such as AdC and alumina) deposited on conducting substrates, fully confirming assessments made back in the early days of XPS.[20,21,22]

Histograms presented in Fig. 2 show the distribution of (a) $E_B^F$, and (b) $E_B^{VL}$ values for the C 1s peak of AdC, thus reflecting differences between the conventional FL referencing and the VL referencing that involves the measurement of the sample work function. In the latter case, the standard deviation is only 0.11 eV, i.e., less than the experimental resolution (~0.3 eV determined mostly by the X-ray width).

It has been suggested that the AdC C 1s peak shifts of the type illustrated in Fig. 1 are caused by the differential charging in the oxide layers that grown on the surface following longer storage in air.[28] Several experimental facts disprove this hypothesis. First, C 1s shifts by 2.5 eV also for samples stored for 14 h in UHV after oxides were removed by Ar etching.[29,30] Secondly, C 1s shifts to higher BE are expected if the differential charging in the oxide should be the reason. However, for many samples the C 1s peaks are at lower binding energy than the "expected" 284.6-285.0 eV (see Fig. 1): for AdC on MoN, VN or WN (all with native oxides) the C 1s peak is found at 284.08, 284.15, and 284.22 eV, respectively. Such low $E_B^F$ values, which directly disprove the differential charging hypothesis, are nicely explained within the framework of the VL alignment by very high work functions of these samples: 5.35, 5.16, and 5.23 eV. Furthermore, in the control experiment specifically designed to verify the "differential charging" hypothesis the C 1s from AdC on Al foil was consistently found at 286.4 eV, irrespective of the native oxide thickness.[31] The latter ranged from 0.7 to 4.7 nm, while the inelastic mean free path for Al 2p electrons in Al was 2.8 nm, i.e., 4 times longer than the minimum Al oxide thickness. It is unbelievable that such thin oxide could develop the positive potential of 1.6 V necessary to move the C 1s peak from the "standard" position of 284.8 eV. Finally, the $E_B^F$ of the C 1s peak shows excellent correlation to the sample work function (cf. Fig. 1), which can not be explained by differential charging.

The VL alignment at the AdC/sample interface contradicts the implicit assumption behind the C 1s method, i.e., that the C 1s peak would appear at the constant BE, which is equivalent of claiming the FL alignment at the AdC/sample interface. The latter would, however, require significant charge transfer across the AdC/sample interface,[25,32] which is contradicted by all experimental evidence about the AdC properties and behavior. The direct implication of the VL alignment is that the core levels from AdC shift as the sample work function changes.[20,21,22,33] Hence, no meaningful referencing to the C 1s peak of AdC is possible without the simultaneous measurement of $\phi_{SA}$. This is true irrespective of the experimental setup used: with samples grounded or isolated from spectrometer (with flood gun controlling the surface potential).

In fact, the VL alignment at the AdC/sample interface should not come as a surprise as AdC layers are not an inherent part of the analyzed sample. They are believed to be composed of aliphatic hydrocarbons and carbooxide fragments,[34,35] which typically do not form strong bonds to the surface and are easily removed in UHV by gentle heating.[36] Thus, they belong to the same category as weakly interacting thin films of saturated hydrocarbons prepared by spin-coating methods, for which VL alignment is well-proven.[37,38,39,40] Such contacts often remain within the Schottky-Mott limit, where the electronic levels of the adsorbate are determined by the work function of the substrate.[41] VL alignment for insulators have also been confirmed for model catalysts system consisting of Ni, Pd, and Pt on single crystal of α-alumina,[42] adsorbates on metal surfaces with sub-monolayer or monolayer coverage,[43,44] and more recently for alumina films on conducting substrates.[27]

The specific value of 289.58±0.12 eV found for the C 1s peak of AdC referenced to VL compares well to values reported for the saturated hydrocarbon molecules. For example, Pireaux et al. reported 290.76 eV for C 1s from alkanes,[45] Karlsen et al. showed that the C 1s peak position varies from 290.71 eV for methane to 290.22 for octane.[46] Somewhat lower values were published for aromatic C moieties such as benzene (290.2 eV),[47] aniline (289.70 eV),[48] or phenol (290.2 eV).[48] All these values are 0.6 to 1.2 eV higher than the 289.58±0.12 eV found for AdC. The latter value is, however, affected by the solid dielectric polarization energy and inter-molecular relaxation. The sum of both contributions, that are absent in the gas phase, can lower the measured BE by up to 1.5 eV.[49]

In summary, the previous finding of the VL alignment at the AdC/sample interface is fully confirmed for a broader sample set comprising 360 thin-film specimens spanning a wide range of material systems such as metals, nitrides, carbides, borides, oxides, carbonitrides, and oxynitrides. The very close correlation between the C 1s position $E_B^F$ and the sample work function $\phi_{SA}$ is confirmed to a better degree than in the previous paper. The variation in the C 1s peak position is 2.89 eV, which is seven times more than the range specified in ISO guidelines, and more than many chemical shifts. Several experimental observations disprove the hypothesis that the C 1s shifts are caused by differential charging. The direct implication of the VL alignment is that no meaningful referencing to the C 1s peak of AdC is possible without the simultaneous measurement of $\phi_{SA}$. This conclusion applies irrespective of whether analyzes are conducted with samples grounded or intentionally isolated from the spectrometer (with flood gun controlling the surface potential).

The author most gratefully acknowledges the following researchers for making samples available: Babak Bakhit, Erik Lewin, Arnaud Le Febvrier, Ching-Lien Hsiao, Barbara Osinger,

Vladyslav Rogoz, Bartosz Wicher, Xiao Li, and Witold Gulbinski. The financial support from the Swedish Energy Agency under projects 51201-1 and P2023-00784, the Åforsk Foundation Grant 22-4, the Olle Engkvist Foundation grant 222-0053, the Swedish Government Strategic Research Area in Materials Science on Advanced Functional Materials at Linköping University (Faculty Grant SFO-Mat-LiU No. 2009-00971), and the Competence Center *Functional Nanoscale Materials* (FunMat-II) VINNOVA grant 2022-03071 is acknowledged.

## Supplementary material

List of all samples included in the study along with measured binding energy values of the C-C/C-H component of the C 1s spectra of adventitious carbon referenced to the Fermi level $E_B^F$ and to the vacuum level $E_B^V$, as well as the sample work function $\phi_{SA}$ assessed by UPS.

# Supplementary file

## Binding energy referencing in X-ray photoelectron spectroscopy: expanded data set confirms that adventitious carbon aligns to the vacuum level

Grzegorz Greczynski

*Thin Film Physics Division, Department of Physics (IFM), Linköping University,*
*SE-581 83 Linköping, Sweden*

List of samples used in the analyses of C 1s peak position from adventitious carbon (AdC) and the work function measurement by UPS. If not otherwise indicated samples are in the form of thin films deposited by magnetron sputtering on Si(001) substrates. AdC is predominantly from air exposure except for a few cases where samples were first etched and then left overnight in the UHV chamber (as marked below).

| | Sample | $E_B^F$ [eV] | $\phi_{SA}$ [eV] | $E_B^V$ [eV] | comments |
|---|---|---|---|---|---|
| 1 | Al | 286.11 | 3.43 | 289.54 | |
| 2 | Si | 285.51 | 4.01 | 289.52 | |
| 3 | Ti | 285.29 | 4.15 | 289.44 | |
| 4 | V | 284.52 | 5.10 | 289.62 | |
| 5 | Cr | 284.59 | 4.79 | 289.38 | |
| 6 | Y | 286.41 | 3.26 | 289.67 | |
| 7 | Zr | 286.12 | 3.56 | 289.68 | |
| 8 | Nb | 285.37 | 4.37 | 289.74 | |
| 9 | Hf | 285.31 | 4.29 | 289.60 | |
| 10 | Ta | 285.73 | 3.98 | 289.71 | |
| 11 | W | 285.18 | 4.56 | 289.74 | |
| 12 | Au | 284.37 | 5.17 | 289.54 | Foil |
| 13 | Ag | 284.82 | 4.89 | 289.71 | Foil |
| 14 | Sc | 285.97 | 3.43 | 289.40 | substrate: $Al_2O_3$(111) |
| 15 | Mn | 285.21 | 4.31 | 289.52 | |
| 16 | Mo | 284.84 | 4.70 | 289.54 | substrate: $Al_2O_3$(111) |
| 17 | Ni | 285.19 | 4.18 | 289.37 | |
| 18 | Cu | 285.21 | 4.23 | 289.44 | foil |
| 19 | Mg | 286.00 | 3.46 | 289.46 | Venting temperature $T_v$ =400 $^oC$ |
| 20 | Al | 286.23 | 3.14 | 289.37 | foil |
| 21 | Fe | 285.67 | 3.82 | 289.49 | Foil |
| 22 | Mg | 286.22 | 3.25 | 289.47 | $T_v$ =29 $^oC$ |
| 23 | Mg | 286.90 | 2.52 | 289.42 | $Ar^+$-etched and exposed to UHV for 14 hours |
| 24 | Pt | 284.54 | 4.96 | 289.5 | |
| 25 | Co | 284.96 | 4.37 | 289.33 | |
| 26 | Cu18Mo82 | 284.66 | 4.85 | 289.51 | |
| 27 | Cu30Mo70 | 285 | 4.6 | 289.6 | |
| 28 | Cu35Mo65 | 284.95 | 4.46 | 289.41 | |
| 29 | Cu41Mo59 | 285.03 | 4.57 | 289.6 | |
| 30 | Cu46Mo54 | 285.09 | 4.41 | 289.5 | |
| 31 | Cu58Mo42 | 284.99 | 4.48 | 289.47 | |
| 32 | Cu68Mo32 | 285.26 | 4.34 | 289.6 | |
| 33 | Cu73Mo27 | 285.21 | 4.38 | 289.59 | |
| 34 | Cu77Mo23 | 285.06 | 4.45 | 289.51 | |
| 35 | Cu82Mo18 | 285.04 | 4.47 | 289.51 | |
| 36 | Cu88Mo12 | 285.06 | 4.45 | 289.51 | |
| 37 | Ti8V28Mo53Nb5W26 | 284.75 | 4.9 | 289.65 | |
| 38 | Ti6V33Mo24Nb7W30 | 284.65 | 4.97 | 289.62 | |
| 39 | Ti6V23Mo9Nb24W38 | 284.46 | 5.08 | 289.54 | |
| 40 | Ti4V18Mo23Nb6W49 | 284.35 | 5.09 | 289.44 | |
| 41 | Ti3V23Mo9Nb24W38 | 284.82 | 4.85 | 289.67 | |
| 42 | Fe65Ti13Co8Ni7Nb1 | 284.92 | 4.57 | 289.49 | |
| 43 | MgAl | 286.83 | 2.69 | 289.52 | AZ31 soild |
| 44 | Co23Cr15Ni16Fe46 | 285.25 | 4.26 | 289.51 | |
| 45 | Co22Cr15Ni16Fe47 | 285.34 | 4.42 | 289.76 | |
| 46 | Co23Cr19Ni14Fe44 | 285.51 | 4.15 | 289.66 | |
| 47 | Co24Cr18Ni13Fe45 | 285.55 | 4.12 | 289.67 | |
| 48 | Co24Cr28Ni10Fe38 | 285.43 | 4.17 | 289.6 | |
| 49 | Co24Cr25Ni10Fe41 | 285.34 | 4.42 | 289.76 | |
| 50 | TiN | 284.52 | 4.9 | 289.42 | |
| 51 | VN | 284.15 | 5.16 | 289.31 | |
| 52 | CrN | 284.6 | 4.83 | 289.43 | |
| 53 | ZrN | 285.49 | 4.09 | 289.58 | |
| 54 | NbN | 284.98 | 4.48 | 289.46 | |
| 55 | MoN | 284.08 | 5.35 | 289.43 | |
| 56 | HfN | 285.52 | 4 | 289.52 | |
| 57 | TaN | 285.08 | 4.41 | 289.49 | |
| 58 | WN | 284.22 | 5.23 | 289.45 | |
| 59 | TiN | 284.68 | 4.70 | 289.38 | $T_v$ = 29 $^oC$ |
| 60 | TiN | 284.65 | 4.69 | 289.34 | $T_v$ = 330 $^oC$ |
| 61 | TiN | 284.62 | 4.71 | 289.33 | steel substrate |
| 62 | $Ti_{0.84}Ta_{0.16}N$ | 284.70 | 4.68 | 289.38 | |
| 63 | $Ti_{0.62}Ta_{0.38}N$ | 284.72 | 4.65 | 289.37 | |
| 64 | $Ti_{0.39}Ta_{0.61}N$ | 284.92 | 4.47 | 289.39 | |
| 65 | $Ti_{0.21}Ta_{0.79}N$ | 285.09 | 4.29 | 289.38 | |
| 66 | $Ti_{0.14}Ta_{0.86}N$ | 285.21 | 4.26 | 289.47 | |
| 67 | $Ti_{0.07}Ta_{0.93}N$ | 285.23 | 4.22 | 289.45 | |
| 68 | $Zr_{0.93}Al_{0.07}N$ | 285.46 | 4.18 | 289.64 | |
| 69 | $Zr_{0.66}Al_{0.34}N$ | 285.74 | 4.08 | 289.82 | |
| 70 | $Zr_{0.75}Al_{0.25}N$ | 285.59 | 4.20 | 289.79 | |
| 71 | $Zr_{0.37}Al_{0.63}N$ | 285.82 | 3.79 | 289.61 | |
| 72 | ZrN | 285.45 | 4.21 | 289.66 | steel substrate |
| 73 | $Zr_{0.73}Al_{0.27}N$ | 285.42 | 4.24 | 289.66 | |
| 74 | $Zr_{0.84}Al_{0.16}N$ | 285.41 | 4.21 | 289.62 | |
| 75 | $Cr_{0.82}Al_{0.18}N$ | 284.77 | 4.70 | 289.47 | |
| 76 | $Cr_{0.71}Al_{0.29}N$ | 284.75 | 4.72 | 289.47 | |
| 77 | $Cr_{0.61}Al_{0.39}N$ | 284.83 | 4.63 | 289.46 | |
| 78 | $Cr_{0.54}Al_{0.46}N$ | 284.92 | 4.56 | 289.48 | |
| 79 | $Cr_{0.45}Al_{0.55}N$ | 284.99 | 4.51 | 289.50 | |
| 80 | $Cr_{0.34}Al_{0.66}N$ | 285.04 | 4.45 | 289.49 | |
| 81 | $Cr_{0.23}Al_{0.77}N$ | 285.04 | 4.45 | 289.49 | |
| 82 | $Cr_{0.15}Al_{0.85}N$ | 285.14 | 4.37 | 289.51 | |
| 83 | $Ti_{0.84}Al_{0.16}N$ | 284.91 | 4.64 | 289.55 | |
| 84 | $Ti_{0.75}Al_{0.25}N$ | 284.80 | 4.82 | 289.62 | |
| 85 | $Ti_{0.65}Al_{0.35}N$ | 284.93 | 4.70 | 289.63 | |
| 86 | $Ti_{0.50}Al_{0.50}N$ | 285.08 | 4.69 | 289.77 | |
| 87 | $Ti_{0.46}Al_{0.54}N$ | 285.20 | 4.53 | 289.73 | |
| 88 | $Ti_{0.41}Al_{0.59}N$ | 285.04 | 4.65 | 289.69 | |
| 89 | $Ti_{0.34}Al_{0.66}N$ | 285.22 | 4.61 | 289.83 | |
| 90 | $Ti_{0.24}Al_{0.76}N$ | 285.41 | 4.46 | 289.87 | |
| 91 | Ti92W8N | 284.9 | 4.59 | 289.49 | |
| 92 | Ti85Al7W8N | 284.91 | 4.56 | 289.47 | |
| 93 | Ti68Al24W8N | 285.08 | 4.51 | 289.59 | |
| 94 | Ti62Al30W8N | 284.79 | 4.62 | 289.41 | |
| 95 | Ti52Al40W8N | 285 | 4.52 | 289.52 | |
| 96 | Ti43Al49W8N | 284.89 | 4.59 | 289.48 | |
| 97 | Ti36Al56W8N | 284.92 | 4.49 | 289.41 | |
| 98 | Ti28Al64W8N | 285.06 | 4.4 | 289.46 | |
| 99 | Ti6V22Mo37Nb5W30N | 284.4 | 5.05 | 289.45 | |
| 100 | V17Mo43W40N | 284.35 | 5.18 | 289.53 | |
| 101 | Ti4V17Mo27Nb5W47N | 284.4 | 5.18 | 289.58 | |
| 102 | V14Mo27W59N | 284.38 | 5.15 | 289.53 | |
| 103 | V11Mo20W69N | 284.33 | 5.15 | 289.48 | |
| 104 | V4Mo17W79N | 284.6 | 4.95 | 289.55 | |
| 105 | V22Mo29W49N | 284.58 | 5.04 | 289.62 | |
| 106 | V31Mo27W42N | 284.32 | 5.16 | 289.48 | |
| 107 | Ti8V28Mo28Nb8W27N | 284.48 | 5.14 | 289.62 | |
| 108 | Hf89Si11N | 285.3 | 4.28 | 289.58 | |
| 109 | Hf86Si14N | 285.42 | 4.33 | 289.75 | |
| 110 | Hf82Si18N | 285.56 | 4.24 | 289.8 | |
| 111 | Hf68Si32N | 286 | 3.58 | 289.58 | |
| 112 | Ti0.89Si0.11N | 284.85 | 4.7 | 289.55 | |
| 113 | Ti0.77Si0.23N | 284.88 | 4.77 | 289.65 | |
| 114 | Ti0.62Si0.38N | 284.72 | 4.8 | 289.52 | |
| 115 | Ti0.54Si0.46N | 284.87 | 4.71 | 289.58 | |
| 116 | Ti0.47Si0.53N | 284.95 | 4.65 | 289.6 | |
| 117 | Ti0.36Si0.64N | 285 | 4.67 | 289.67 | |
| 118 | Ti0.23Si0.77N | 284.93 | 4.62 | 289.55 | |
| 119 | Ti0.11Si089N | 285.04 | 4.45 | 289.49 | |
| 120 | Mo72V27N | 284.56 | 4.99 | 289.55 | |
| 121 | Mo60V40N | 284.53 | 4.97 | 289.5 | |
| 122 | Mo65V35N | 284.48 | 5.03 | 289.51 | |
| 123 | Mo50V50N | 284.39 | 5.04 | 289.43 | |
| 124 | Mo52V48N | 284.46 | 4.92 | 289.38 | |
| 125 | Mo62V38N | 284.52 | 4.96 | 289.48 | |
| 126 | Mo51V49N | 284.54 | 5 | 289.54 | |
| 127 | Mo50V50N | 284.42 | 5.02 | 289.44 | |
| 128 | Mo47V53N | 284.52 | 4.98 | 289.5 | |
| 129 | Mo45V55N | 284.48 | 5.02 | 289.5 | |
| 130 | V84Al16N | 284.68 | 4.82 | 289.5 | |
| 131 | V77Al23N | 284.78 | 4.67 | 289.45 | |
| 132 | V67Al33N | 284.82 | 4.68 | 289.5 | |
| 133 | V61Al39N | 284.76 | 4.67 | 289.43 | |
| 134 | V55Al45N | 284.81 | 4.64 | 289.45 | |
| 135 | V50Al50N | 285.04 | 4.56 | 289.6 | |
| 136 | V45Al55N | 285 | 4.5 | 289.5 | |
| 137 | V40Al60N | 284.95 | 4.62 | 289.57 | |
| 138 | V34Al66N | 284.95 | 4.62 | 289.57 | |
| 139 | V30Al70N | 284.96 | 4.63 | 289.59 | |
| 140 | V26Al74N | 284.85 | 4.72 | 289.57 | |
| 141 | V20Al80N | 284.78 | 4.71 | 289.49 | |
| 142 | V16Al84N | 284.74 | 4.75 | 289.49 | |
| 143 | Cr99N1 | 284.77 | 4.76 | 289.53 | |
| 144 | CR95N5 | 284.72 | 4.89 | 289.61 | |
| 145 | Cr89N11 | 284.68 | 4.84 | 289.52 | |
| 146 | Cr85N15 | 284.63 | 4.86 | 289.49 | |
| 147 | Cr78N22 | 284.6 | 4.86 | 289.46 | |
| 148 | Cr75N25 | 284.56 | 4.87 | 289.43 | |
| 149 | Cr73N27 | 284.62 | 4.87 | 289.49 | |
| 150 | Cr64N36 | 284.54 | 4.92 | 289.46 | |
| 151 | Cr58N42 | 284.54 | 4.91 | 289.45 | |
| 152 | Cr56N44 | 284.64 | 4.81 | 289.45 | |
| 153 | Cr53N47 | 284.5 | 4.94 | 289.44 | |
| 154 | Cr46N54 | 284.56 | 4.91 | 289.47 | |
| 155 | $TiB_2$ | 285.40 | 4.47 | 289.97 | |
| 156 | $ZrB_2$ | 285.59 | 4.30 | 289.89 | |
| 157 | $Al_{0.10}Mg_{0.07}B_{0.83}$ | 286.29 | 3.34 | 289.63 | |
| 158 | $Al_{0.07}Mg_{0.08}B_{0.85}$ | 286.18 | 3.49 | 289.67 | |
| 159 | $Al_{0.09}Mg_{0.13}B_{0.78}$ | 286.18 | 3.44 | 289.62 | |
| 160 | $Al_{0.08}Mg_{0.08}B_{0.84}$ | 286.07 | 3.53 | 289.60 | |
| 161 | $Al_{0.07}Mg_{0.03}B_{0.90}$ | 286.18 | 3.36 | 289.54 | |
| 162 | $Al_{0.07}Mg_{0.06}B_{0.87}$ | 286.34 | 3.36 | 289.70 | |
| 163 | $Al_{0.08}Mg_{0.16}B_{0.76}$ | 286.11 | 3.50 | 289.61 | |
| 164 | Ti0.64Al0.36B1.93 | 285.84 | 3.72 | 289.56 | |
| 165 | Ti0.51Al0.49B1.97 | 285.68 | 3.95 | 289.63 | |
| 166 | Ti0.42Al0.58B1.87 | 285.78 | 3.86 | 289.64 | |
| 167 | Ti0.38Al0.62B1.92 | 285.7 | 3.87 | 289.57 | |
| 168 | Ti0.33Al0.67B1.83 | 285.82 | 3.79 | 289.61 | |
| 169 | Ti0.26Al0.74B1.82 | 285.78 | 3.77 | 289.55 | |
| 170 | Ti0.6Al0.4B2.03 | 285.64 | 3.94 | 289.58 | |
| 171 | Ti0.26Al0.74B1.86 | 285.77 | 3.78 | 289.55 | |
| 172 | Ti0.24Al0.76B1.81 | 285.81 | 3.72 | 289.53 | |
| 173 | AlB11.5 | 285.09 | 4.5 | 289.59 | |
| 174 | AlB3.27 | 285.7 | 3.88 | 289.58 | |
| 175 | AlB2.75 | 285.66 | 3.91 | 289.57 | |
| 176 | AlB2.5 | 285.55 | 4.02 | 289.57 | |
| 177 | AlB2.3 | 285.53 | 4.09 | 289.62 | |
| 178 | AlB2.2 | 285.94 | 3.63 | 289.57 | |
| 179 | AlB1.83 | 285.5 | 4.1 | 289.60 | |
| 180 | Ti85Si15B2.44 | 285.59 | 4.08 | 289.67 | |
| 181 | Ti75Si25B2.07 | 285.67 | 4.02 | 289.69 | |
| 182 | Ti69Si31B1.77 | 285.69 | 4.02 | 289.71 | |
| 183 | Ti86Si14B2.39 | 285.62 | 4.23 | 289.85 | |
| 184 | Ti77Si23B2.06 | 285.56 | 4.17 | 289.73 | |
| 185 | Ti72Si28B1.84 | 285.51 | 4.13 | 289.64 | |
| 186 | Ti62Si38B1.55 | 285.64 | 4.16 | 289.8 | |
| 187 | HfB2 | 285.79 | 3.85 | 289.64 | |
| 188 | W2B5 | 285.28 | 4.35 | 289.63 | |
| 189 | TiWB2.1 | 285.28 | 4.35 | 289.63 | |
| 190 | TiWB2.3 | 285.31 | 4.32 | 289.63 | |
| 191 | VB2 | 284.73 | 4.74 | 289.47 | |
| 192 | VB2 | 285.32 | 4.41 | 289.73 | $Ar^+$-etched and exposed to UHV for 14 hours |
| 193 | TaB2 | 285.5 | 4 | 289.5 | |
| 194 | CrB1.9 | 285.06 | 4.51 | 289.57 | |
| 195 | CrB1.8 | 284.96 | 4.6 | 289.56 | |
| 196 | CrB1.75 | 284.92 | 4.7 | 289.62 | |
| 197 | Zr77Cr23B1.5 | 285.5 | 3.85 | 289.35 | |
| 198 | Zr71Cr29B1.42 | 285.7 | 3.72 | 289.42 | |
| 199 | Zr68Cr32B1.38 | 285.53 | 3.84 | 289.37 | |
| 200 | Zr64Cr36B1.3 | 285.51 | 3.88 | 289.39 | |
| 201 | Zr56Cr44B1.11 | 285.68 | 3.79 | 289.47 | |
| 202 | Zr71Cr29B1.42 | 285.49 | 3.94 | 289.43 | $Ar^+$-etched and exposed to UHV for 14 hours |
| 203 | Zr68Cr32B1.38 | 285.56 | 3.87 | 289.43 | $Ar^+$-etched and exposed to UHV for 14 hours |
| 204 | Zr64Cr36B1.3 | 285.63 | 3.81 | 289.44 | $Ar^+$-etched and exposed to UHV for 14 hours |
| 205 | Zr56Cr44B1.11 | 285.67 | 3.81 | 289.48 | $Ar^+$-etched and exposed to UHV for 14 hours |
| 206 | Zr52V48B2 | 285.18 | 4.3 | 289.48 | |
| 207 | Z46rV54B2 | 285.22 | 4.16 | 289.38 | |
| 208 | Zr0.9Nb0.1B2 | 285.66 | 3.84 | 289.5 | |
| 209 | Zr0.8Nb0.2B2 | 285.83 | 3.72 | 289.55 | |
| 210 | Zr0.7Nb0.3B2 | 285.8 | 3.72 | 289.52 | |
| 211 | Zr0.9Ta0.1B2.1 | 285.65 | 3.84 | 289.49 | |
| 212 | Zr0.8Ta0.2B1.8 | 285.72 | 3.81 | 289.53 | |
| 213 | Zr0.7Ta0.3B1.5 | 285.7 | 3.83 | 289.53 | |
| 214 | ZrB2.3 | 285.7 | 3.79 | 289.49 | |
| 215 | Ti15Al28B57 | 285.98 | 3.53 | 289.51 | |
| 216 | Ti14Al29B57 | 285.48 | 4.15 | 289.63 | |
| 217 | Ti12Al28B60 | 285.41 | 4.07 | 289.48 | |
| 218 | Ti16Al30B54 | 285.77 | 3.77 | 289.54 | |
| 219 | Ti27Al11B62 | 285.44 | 4.06 | 289.5 | |
| 220 | Ti17Al27B57 | 285.71 | 3.83 | 289.54 | |
| 221 | Ti18Al26B56 | 285.68 | 3.86 | 289.54 | |
| 222 | Ti19Al23B58 | 285.67 | 3.89 | 289.56 | |
| 223 | Ti17Al28B55 | 285.72 | 3.82 | 289.54 | |
| 224 | Ti18Al24B58 | 285.71 | 3.87 | 289.58 | |
| 225 | Ti19Al23B58 | 285.67 | 3.92 | 289.59 | |
| 226 | Ti19Al24B57 | 285.69 | 3.88 | 289.57 | |
| 227 | Ti15Al30B55 | 286.13 | 3.45 | 289.58 | |
| 228 | Ti14Al36B50 | 285.78 | 3.78 | 289.56 | |
| 229 | Ti27Al19B54 | 285.76 | 3.89 | 289.65 | |
| 230 | Hf17Ti22B61 | 285.63 | 3.88 | 289.51 | |
| 231 | Hf24Ti21B55 | 285.52 | 3.94 | 289.46 | |
| 232 | Hf10Ti30B60 | 285.42 | 4.14 | 289.56 | |
| 233 | Hf15Ti29B56 | 285.42 | 4.1 | 289.52 | |
| 234 | Hf26Ti22B52 | 285.72 | 4 | 289.72 | |
| 235 | Hf39Ti7B54 | 285.83 | 3.8 | 289.63 | |
| 236 | Hf20Ti22B58 | 285.59 | 4.02 | 289.61 | |
| 237 | Hf25Ti22B53 | 285.72 | 3.89 | 289.61 | |
| 238 | Hf42Ti5B53 | 285.85 | 3.75 | 289.6 | |
| 239 | TiBx | 285.59 | 3.95 | 289.54 | |
| 240 | Ti77Hf33B1.7 | 285.51 | 4.05 | 289.56 | |
| 241 | TiHfB | 285.53 | 4.06 | 289.59 | |
| 242 | TiHfB | 285.64 | 4.05 | 289.69 | |
| 243 | TiBx | 285.48 | 4.02 | 289.5 | |
| 244 | Hf46B54 | 285.78 | 3.86 | 289.64 | |
| 245 | Hf47B53 | 285.92 | 3.55 | 289.47 | |
| 246 | HfB | 285.61 | 3.94 | 289.55 | |
| 247 | Hf51B49 | 285.61 | 3.94 | 289.55 | |
| 248 | Hf49B51 | 285.67 | 3.9 | 289.57 | |
| 249 | Hf52B48 | 285.69 | 3.93 | 289.62 | |
| 250 | Hf55B45 | 285.78 | 3.89 | 289.67 | |
| 251 | TiC | 284.92 | 4.76 | 289.68 | |
| 252 | VC | 284.67 | 4.97 | 289.64 | |
| 253 | CrC | 284.86 | 4.83 | 289.69 | |
| 254 | NbC | 284.98 | 4.82 | 289.80 | |
| 255 | MoC | 284.83 | 4.93 | 289.76 | |
| 256 | Ni74Cr22C4 | 285.31 | 4.13 | 289.44 | |
| 257 | Ni73Cr21C6 | 285.23 | 4.33 | 289.56 | |
| 258 | Ni69Cr20C11 | 285.18 | 4.27 | 289.45 | |
| 259 | Ni57Cr22C21 | 285.08 | 4.23 | 289.31 | |
| 260 | Ni52Cr21C27 | 285 | 4.34 | 289.34 | |
| 261 | Ni36Cr16C48 | 285.11 | 4.27 | 289.38 | |
| 262 | Ni26Cr11C63 | 285.02 | 4.28 | 289.3 | |
| 263 | ZrC | 285.14 | 4.48 | 289.62 | |
| 264 | HfC | 285.44 | 4.22 | 289.66 | |
| 265 | TaC | 285.14 | 4.52 | 289.66 | |
| 266 | WC | 285.09 | 4.58 | 289.67 | |
| 267 | WC1.2 | 284.89 | 4.62 | 289.51 | |
| 268 | WC1.5 | 285.29 | 4.44 | 289.73 | |
| 269 | WC1.5 | 285.12 | 4.69 | 289.81 | $Ar^+$-etched and exposed to UHV for 14 hours |
| 270 | Ti41C59 | 284.84 | 4.51 | 289.35 | |
| 271 | Ti43C57 | 284.75 | 4.58 | 289.33 | |
| 272 | Ti47C53 | 285.03 | 4.56 | 289.59 | |
| 273 | Ti54C46 | 285.06 | 4.34 | 289.4 | |
| 274 | Ti51C49 | 285.04 | 4.52 | 289.56 | |
| 275 | Ti45C55 | 284.96 | 4.55 | 289.51 | |
| 276 | Ti42C58 | 284.94 | 4.59 | 289.53 | |
| 277 | Ti41C59 | 284.94 | 4.58 | 289.52 | |
| 278 | Ti55C45 | 284.79 | 4.74 | 289.53 | |
| 279 | Ni68Mo22C10 | 285.26 | 4.18 | 289.44 | |
| 280 | Ni69Mo21C10 | 285.05 | 4.43 | 289.48 | |
| 281 | Ni70Mo20C10 | 285.04 | 4.43 | 289.47 | |
| 282 | Ni62Mo22C16 | 285.08 | 4.42 | 289.5 | |
| 283 | Ni57Mo19C24 | 285.09 | 4.36 | 289.45 | |
| 284 | Ni43Mo14C43 | 284.94 | 4.39 | 289.33 | |
| 285 | Ni28Mo9C63 | 284.81 | 4.42 | 289.23 | |
| 286 | Ni19Mo5C76 | 284.86 | 4.46 | 289.32 | |
| 287 | ZrTi2AlC | 285.63 | 3.87 | 289.5 | |
| 288 | V9Al28C63 | 285.28 | 4.17 | 289.45 | |
| 289 | V10Al32C58 | 285.31 | 4.15 | 289.46 | |
| 290 | V12Al33C55 | 285.28 | 4.16 | 289.44 | |
| 291 | V11Al35C54 | 285.27 | 4.17 | 289.44 | |
| 292 | V12Al34C54 | 285.31 | 4.13 | 289.44 | |
| 293 | V10Al36C54 | 285.42 | 3.99 | 289.41 | |
| 294 | HfC | 286.34 | 3.06 | 289.4 | |
| 295 | ZrC | 285.4 | 4.05 | 289.45 | |
| 296 | WC | 285.06 | 4.39 | 289.45 | |
| 297 | NbC | 284.66 | 4.73 | 289.39 | |
| 298 | VC | 284.73 | 4.81 | 289.54 | |
| 299 | NbC | 285.77 | 3.91 | 289.68 | |
| 300 | TaC | 285.15 | 4.24 | 289.39 | |
| 301 | MoC | 284.95 | 4.37 | 289.32 | |
| 302 | $V_{0.18}Al_{0.31}O_{0.03}N_{0.48}$ | 284.90 | 4.75 | 289.65 | |
| 303 | $V_{0.16}Al_{0.34}O_{0.03}N_{0.47}$ | 284.92 | 4.78 | 289.70 | |
| 304 | $V_{0.16}Al_{0.34}O_{0.04}N_{0.46}$ | 284.88 | 4.88 | 289.76 | |
| 305 | $V_{0.17}Al_{0.33}O_{0.07}N_{0.43}$ | 285.17 | 4.49 | 289.66 | |
| 306 | $V_{0.19}Al_{0.33}O_{0.11}N_{0.38}$ | 285.05 | 4.63 | 289.68 | |
| 307 | $V_{0.20}Al_{0.33}O_{0.17}N_{0.30}$ | 284.80 | 4.88 | 289.68 | |
| 308 | $V_{0.22}Al_{0.30}O_{0.25}N_{0.23}$ | 284.70 | 4.97 | 289.67 | |
| 309 | $V_{0.24}Al_{0.28}O_{0.30}N_{0.18}$ | 284.67 | 5.05 | 289.72 | |
| 310 | $V_{0.19}Al_{0.35}O_{0.38}N_{0.08}$ | 284.61 | 5.06 | 289.67 | |
| 311 | Cr92C6N2 | 284.75 | 4.77 | 289.52 | |
| 312 | Cr72C20N8 | 284.83 | 4.82 | 289.65 | |
| 313 | Cr64C25N11 | 284.87 | 4.81 | 289.68 | |
| 314 | Cr62C32N6 | 284.84 | 4.79 | 289.63 | |
| 315 | $Ta_2O_5$ | 286.10 | 3.48 | 289.58 | |
| 316 | $TiO_2$ | 285.30 | 4.22 | 289.52 | |
| 317 | Al2O3(10nm)/Al | 286.97 | 2.6 | 289.57 | |
| 318 | Al2O3(10nm)/W | 286.34 | 3.23 | 289.57 | |
| 319 | Al2O3(10nm)/TiN | 286.65 | 3.02 | 289.67 | |
| 320 | Al2O3(10nm)/HfN | 287.28 | 2.2 | 289.48 | |
| 321 | Al2O3(10nm)/Si | 286.76 | 2.87 | 289.63 | |
| 322 | Al2O3(10nm)/VN | 286.38 | 3.4 | 289.78 | |
| 323 | Al2O3(10nm)/V | 286.37 | 3.32 | 289.69 | |
| 324 | Al2O3(10nm)/WN | 286.33 | 3.26 | 289.59 | |
| 325 | Al2O3(10nm)/Zr | 286.93 | 2.7 | 289.63 | |
| 326 | Al2O3(2nm)/Al | 286.5 | 2.94 | 289.44 | |
| 327 | Al2O3(2nm)/W | 285.45 | 3.88 | 289.33 | |
| 328 | Al2O3(2nm)/MoN | 285.13 | 4.41 | 289.54 | |
| 329 | Al2O3(2nm)/TiN | 285.57 | 4.12 | 289.69 | |
| 330 | Al2O3(2nm)/HfN | 286.78 | 3.02 | 289.8 | |
| 331 | Al2O3(2nm)/Si | 286.17 | 3.37 | 289.54 | |
| 332 | Al2O3(2nm)/VN | 285.04 | 4.48 | 289.52 | |
| 333 | Al2O3(2nm)/V | 285.18 | 4.29 | 289.47 | |
| 334 | Al2O3(2nm)/WN | 285.66 | 3.9 | 289.56 | |
| 335 | Al2O3(2nm)/Zr | 286.3 | 3.21 | 289.51 | |
| 336 | Al2O3(4.8nm)/Al | 286.17 | 3.3 | 289.47 | |
| 337 | Al2O3(5.8nm)/Al | 286.39 | 3.17 | 289.56 | |
| 338 | Al2O3(3.7nm)/Al | 286.06 | 3.29 | 289.35 | |
| 339 | Al2O3(7.3nm)/Al | 286.79 | 2.87 | 289.66 | |
| 340 | Al2O3(11.8nm)/Al | 286.95 | 2.76 | 289.71 | |
| 341 | Al2O3(1.1nm)/W | 285.3 | 4.12 | 289.42 | |
| 342 | Al2O3(4nm)/W | 285.74 | 3.66 | 289.4 | |
| 343 | Al2O3(6nm)/W | 285.83 | 3.58 | 289.41 | |
| 344 | Al2O3(8nm)/W | 286.09 | 3.47 | 289.56 | |
| 345 | Al2O3(12nm)/W | 286.46 | 3.3 | 289.76 | |
| 346 | SiO2(1nm)/W | 285.4774 | 4.04 | 289.52 | |
| 347 | SiO2(2nm)/W | 285.659 | 3.94 | 289.6 | |
| 348 | SiO2(4nm)/W | 285.7508 | 3.79 | 289.54 | |
| 349 | SiO2(6nm)/W | 285.8166 | 3.77 | 289.59 | |
| 350 | SiO2(8nm)/W | 285.8671 | 3.70 | 289.57 | |
| 351 | SiO2(10nm)/W | 285.8933 | 3.73 | 289.62 | |
| 352 | SiO2(12nm)/W | 285.9381 | 3.67 | 289.61 | |
| 353 | SiO2(15nm)/W | 285.9889 | 3.70 | 289.69 | |
| 354 | HfO2(1nm)/W | 285.3 | 4.14 | 289.44 | |
| 355 | HfO2(2nm)/W | 285.4 | 4.19 | 289.59 | |
| 356 | HfO2(4nm)/W | 285.7 | 3.88 | 289.58 | |
| 357 | HfO2(6nm)/W | 286 | 3.57 | 289.57 | |
| 358 | HfO2(8nm)/W | 286.1 | 3.47 | 289.57 | |
| 359 | HfO2(10nm)/W | 286.2 | 3.4 | 289.6 | |
| 360 | HfO2(12nm)/W | 286.2 | 3.51 | 289.71 | |
| 361 | HfO2(15nm)/W | 286.4 | 3.4 | 289.8 | |
| 362 | CuO(10nm)/W | 285.02 | 4.55 | 289.57 | |
| | | | | | |
| | | | | | |